\documentclass[sigchi-a]{acmart}
\usepackage{booktabs} 
\usepackage{ccicons}  

\copyrightyear{2019} 
\acmYear{2019} 
\setcopyright{rightsretained} 
\acmConference[CHI'19 Extended Abstracts]{CHI Conference on Human Factors in Computing Systems Extended Abstracts}{May 4--9, 2019}{Glasgow, Scotland UK}
\acmBooktitle{CHI Conference on Human Factors in Computing Systems Extended Abstracts (CHI'19 Extended Abstracts), May 4--9, 2019, Glasgow, Scotland UK}
\acmDOI{10.1145/3290607.3312973}
\acmISBN{978-1-4503-5971-9/19/05}

\begin{document}
\title{Older Adults and Voice Interaction: A Pilot Study with Google Home}

\author{Jaros{\l}aw Kowalski}
\affiliation{%
  \institution{National Information Processing Institute}
  \streetaddress{al. Niepodleg{\l}o{\s}ci 188b}
  \postcode{00-608}
  \city{Warsaw}
  \country{Poland}
}
\email{jaroslaw.kowalski@opi.org.pl}

\author{Anna Jaskulska}
\affiliation{%
  \institution{Kobo Association}
  \streetaddress{Nowolipie 25b Str.}
  \postcode{02-008}
  \city{Warsaw}
  \country{Poland}
}
\email{a.jaskulska@kobo.org.pl}

\author{Kinga Skorupska}
\affiliation{%
  \institution{Polish-Japanese Academy of Information Technology}
  \streetaddress{86 Koszykowa Str.}
  \postcode{02-008}
  \city{Warsaw}
  \country{Poland}
}
\email{kinga.skorupska@pja.edu.pl}

\author{Katarzyna Abramczuk \\
  Cezary Biele}
\affiliation{%
  \institution{National Information Processing Institute}
  \streetaddress{al. Niepodleg{\l}o{\s}ci 188b}
  \postcode{00-608}
  \city{Warsaw}
  \country{Poland}}
\email{katarzyna.abramczuk@opi.org.pl}
\email{cezary.biele@opi.org.pl}

\author{Wies\l{}aw Kope\'{c}}
  \orcid{0000-0001-9132-4171}
  \affiliation{%
  \institution{Polish-Japanese Academy of Information Technology}
  \streetaddress{86 Koszykowa Str.}
  \postcode{02-008}
  \city{Warsaw}
  \country{Poland}}
\email{kopec@pja.edu.pl}

\author{Krzysztof Marasek}
  \affiliation{%
  \institution{Polish-Japanese Academy of Information Technology}
  \streetaddress{86 Koszykowa Str.}
  \postcode{02-008}
  \city{Warsaw}
  \country{Poland}}
\email{kmarasek@pja.edu.pl}

\renewcommand{\shortauthors}{J. Kowalski et al.}

\begin{abstract}
In this paper we present the results of an exploratory study examining the potential of voice assistants (VA) for some groups of older adults in the context of Smart Home Technology (SHT). To research the aspect of older adults' interaction with voice user interfaces (VUI) we organized two workshops and gathered insights concerning possible benefits and barriers to the use of VA combined with SHT by older adults. Apart from evaluating the participants' interaction with the devices during the two workshops we also discuss some improvements to the VA interaction paradigm.
\end{abstract}

\keywords{IoT; older adults; smart home; living lab; voice assistant; VUI; chatbots; google home}


\maketitle

\begin{marginfigure}
\centering
   \includegraphics[width=170px]{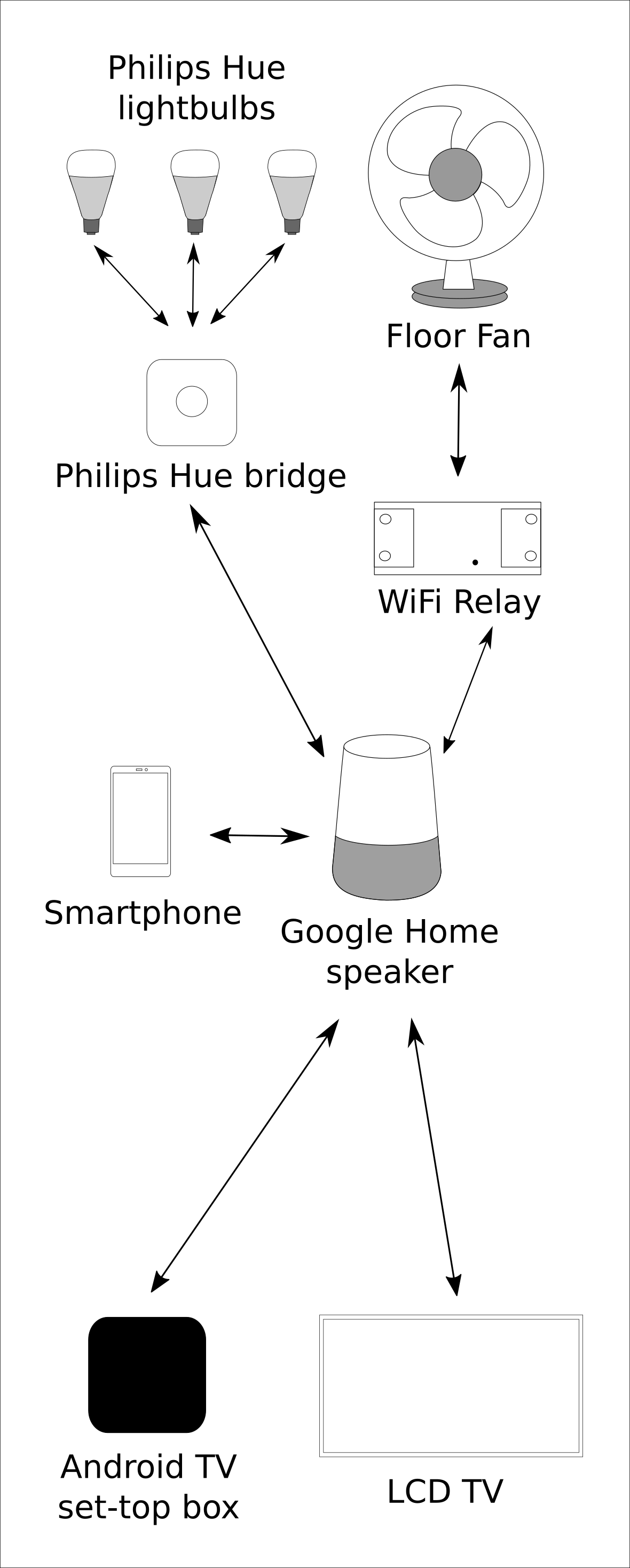}
   \caption{Equipment connection diagram}
   \label{fig:connections}
\end{marginfigure}

\section{Introduction}
Advancements in AI and ML which drive improvements in NLP are making Voice User Interfaces (VUI) increasingly user-friendly and accessible to users with little prior ICT training. This may prove exceptionally useful for some groups of older adults, as it can empower them to actively and comfortably use ICT-enabled solutions on their own. To this end, it is key to discover potential barriers and ways to build on the strengths of this mode of interaction. Such insights could further inform the design of commercial devices, such as SHT with VUIs, to take advantage of the Silver Economy.

To research the older adults' interaction with Voice User Interfaces combined with Smart Home Technology we conducted two focus-group studies, following the same scenario, the results of which are presented in this abstract. First, we discuss the state of the art. Second, we describe the setup of the study, characteristics of our participants and the outline of our scenario. Third, we present the results of our study showcasing the functions our participants discovered and expected. This section is followed by our insights into the benefits and barriers of this technology. Finally, we present our conclusions and discuss the potential future work in this field.

\section{Related work}

Multiple studies on HCI and aging, instead of exploring the aging process and opportunities, focus on health, socialization and technology stereotypes \cite{articleageoldproblem}. Positive studies, are still rare, and one shared feature is that they use older adults' strengths, such as their insights from life experience \cite{Kopeć2018} or language proficiency \cite{skorupska2018smarttv} which increase with age. Moreover, older adults realize the benefits that come with an increased ICT proficiency \cite{aula_learning_2004} especially if it helps them achieve personal goals \cite{djoub_ict_2013}, such as communicating with their loved ones, managing finance, engaging in e-commerce or pursuing their interests \cite{boulton2007ageing, naumanen_practices_2008, von2018influence}.

For this reason, some older adults are enthusiastic about joining LivingLab ICT initiatives \cite{kopec2017living} despite some barriers to their involvement in ICT \cite{sandhu2013ict, kopec2017spiral}. Although there are previous studies on VA usability \cite{Pyae:2018:IUU:3236112.3236130}, also with older adults in the context of SH \cite{Vacher:2015:ECV:2785580.2738047} it is worth to further investigate this problem in multiple cultural contexts in order to suggest how to make Voice Interfaces more accessible \cite{Corbett:2016:ISA:2935334.2935386} in general to facilitate different groups of older adults' interaction with ICT-enhanced solutions.

\section{Methods}
Voice interaction has been implemented in mobile phones and tablets for several years now, but only recently standalone devices like Google Home, Amazon Alexa or Apple HomePod, became available. With their help it is possible to ask queries, perform tasks and control a number of Smart Home devices - and all of these aspects we addressed in our study setup.

\begin{margintable}
\caption{List of equipment}
\label{equipment}
\begin{tabular}{p{2.5cm}p{4cm}}
	\hline
	\textbf{Type} & \textbf{Specs} \\
	\hline
    Speaker with VA & Google Home \\
    \hline
   Smartphone 
   & Xiaomi Mi Max 2 running Android 7.1.1\\
	\hline
   TV set & 42" Samsung LCD \\
   	\hline
   Lights & Three Phillips Hue lightbulbs mounted in a desk lamps, connected to a Phillips Hue bridge\\
   	\hline
    WiFi Relay + Fan & Sonoff Basic WiFi smart switch connected to a 40 cm floor fan\\
    \hline
\end{tabular}
\end{margintable}

We used a system consisting of speaker with a voice assistant agent, smart phone, and additional peripherals such as a TV set, an Android TV Set-Top Box, light controllers, Wi-Fi relays and other. For the purpose of the study all the necessary online accounts were created and preconnected in the Google Home application. The used equipment is listed in Table \ref{equipment}, and the connections are depicted in Figure \ref{fig:connections}.

We invited seven older adults: three female participants and four male participants from our LivingLabs to take part in our study. They were all retired, but remained active in various aspects e.g. as citizens engaged in their local area. All of them lived in Warsaw, Poland, and they were native Polish speakers with different English language proficiency. There was a 25 year age span: the youngest participant was 64 years old and the oldest one was 89, with a mean age of 73.14 (SD=8.64). Based on previous activities in our Living Lab, i.e. introductory interviews and DigComp surveys we can describe this group as very active users, above basic ICT skills. 
We chose to divide the participants into two small affinity groups to give them time to discuss and test various functionalities freely; the scenario for both sessions is presented in Sidebar \ref{bar:sidebar}. The research was based on semi-structured mixed method scenario of group interviews with direct involvement of the participants, and we recorded and transcribed the sessions to create an affinity diagram of key themes and quotes.

\begin{sidebar}

The study scenario consisted of 6 parts:
\begin{enumerate}
  \item discussing situations in which technology can help users at home,
  \item presentation of the Google Home system and voice control capabilities of devices,
  \item independent use of the Google Home system by participants,
  \item collecting opinions about the system, with particular emphasis on advantages, disadvantages and insights on limitations and opportunities,
  \item brainstorming to generate potential new, unaddressed system applications,
  \item collecting opinions about potential threats.
\end{enumerate}

  \caption{An overview of the study scenario}
  \label{bar:sidebar}
\end{sidebar}

\section{Results}

Overall, we observed that our group of older adults was impressed by the range of possibilities of Voice Assistants and how convenient this technology could be for them. One of them said: "This technology is for older adults, 60+, young people have no time to listen to such information: rush, work, they have a phone with everything in it" (P7). Another noted: "it is much simpler, I don't have to devise anything, I just sit there, bored or tired, and say things (...) I just say them and it (VA) does the job" (P5). 

For our group of older adults four key needs emerged:

\begin{enumerate}
\item understanding technology and receiving feedback ((P3) "technology should give us hints: you did something wrong, do this and that" and "younger people already know how everything works and don't understand that for us everything has to be coherent, while they can omit things, even in explaining them");
\item accessible design with low barrier of entry, unlike regular computers ("using a mouse and searching for information is very difficult, as you need multiple repetitions to become a proficient user" and "screens are cluttered, and such user can not focus on one piece of information" (P7));

\begin{margintable}
\caption{List of observed benefits with some improvement suggestions}
\label{benefits}
\begin{tabular}{|p{3cm}|p{3.5cm}|}
	\hline
	\textbf{Benefits} & \textbf{Improvements} \\
	\hline
    \textbf{Intuitive interaction} (via natural language) with no/little prior training required. & More feedback allowing to create a step-by-step guide. \\
    \hline
   \textbf{Voice control} (no motor functions involved on the part of the user). & Understanding more natural utterances, including context and metaphors, as well as tackling and explaining the problem of voice priority to prevent conflicts that may arise.\\
	\hline
   \textbf{Friendly manner} (friendly voice and patience). & Building up on the voice recognition functions to initiate friendly conversations with reminders. \\
   	\hline
   \textbf{No handling of devices} (so, no device that has to be found and turned on). & Solving concerns about the range of effective voice interaction in their home.\\
   	\hline
    \textbf{Granting independence} (the VA can do some things which, especially for people with disabilities, may require assistance). & Ensuring the existence of fail-safes to resolve concerns about the reliability of VIUs in executing commands.\\
    \hline
\end{tabular}
\end{margintable}

\item seamless incorporation into everyday life, as our participants liked the idea of being able to accomplish certain tasks using only speech not only because they might have "difficulties moving" (P1) but also because it is does not disturb their process. They mentioned the advantages of being able to ask for something when they are "just busy doing something else" (P1) or "when my hands are dirty" (P4). They think this could "save a lot of time, walking and searching" (P3). Hence, they think that voice commands "are the future" (P2));
\item control and assurance of security ("the computer does not do what you want it to do, but what you ordered it to do and you never know if they are the same thing" (P2). Another participant expressed the fear that they "are not in control of the autonomous device" (P5). He wondered whether asking technology to turn off the stove "would really work").
\end{enumerate}

In light of the aforementioned needs conversational interaction seems like a very good solution for older adults. First, it allows the user to proceed at their own pace. Such conversational agent will not "complain that it has no time" (P6). Second, it allows the users to renounce the reliance on screens and input devices and manages the information flow. Third, it is based on the use of natural language making it easy, as conversation is the default mode of interaction and it does not require a significant change of habits. Finally, a person can say a command with a specific task in mind without the fear of being distracted or sidetracked. 

In connection with Smart Home devices our group of older adults saw Google Home as a "central unit" (P1) that could make it easier to "operate other devices" (P4), with the TV set being mentioned by several of them, manage various "intelligent home" settings (P1) and even save energy (P6). They especially wanted IoT devices to assist them with tiring household chores, to free up their time for other activities. In terms of barriers, we could notice the concern about context awareness ("Will it tell me if I should take an umbrella with me, or do I have to ask about the weather first?" by P4) and the fear of losing diversity of devices, as not all of them would be compatible (P2). While the issue of privacy was mentioned (P5) it was not a prominent concern. 

Moreover, we found that our group of older adults could naturally identify various already available applications of VA with SHT and accept them as generally useful and empowering. They expected VA to act as an assistant that enables searching for information hands-free or as a memory aid that does not require handling any devices. Other functions which were mentioned consistently were a translator or even a teacher. With little training and encouragement our participants could start using VA's to their satisfaction and empowerment. As one of our participants put it: "if there were training sessions (...) a lot of older people would warm up to it" (P6).

\section{Discussion}

\begin{margintable}
\caption{Summary of observed barriers with some solutions and comments, which may help mitigate them}
\label{barriers}
\begin{tabular}{|p{3cm}|p{3.5cm}|}
	\hline
	\textbf{Barriers} & \textbf{Solutions and comments} \\
	\hline
    \textbf{Time consuming}. & Although this was mentioned, it was not a drawback for older adults as they say that they have the time. \\
    \hline
   \textbf{Lacks sensors and cameras} (which would allow it to better assist with some tasks). & Connecting a camera, to give the user hints (while cooking with a recipe, or to measure things) as well as sensors.\\
 	\hline
   \textbf{Lack of a screen to give feedback and context}. & Introduction of companion screens to see context, status or key information searched for, as it is hard to store it all in memory. \\
   	\hline
   \textbf{Need to have compatible devices}: fear of losing diversity and individuality). & Working towards compatibility between manufacturers.
   \\ 
   	\hline
    \textbf{Fear of malfunction }(something may not turn off, even if the signal was sent). & Making clear what backup security measures are in place
May go away with more exposure.
\\
    \hline
    \textbf{Fear of too much reliance} (afraid of a possible loss of creativity, and lack of mental and physical exercise). & VA could also serve as an assistant, verifying their cognitive health and reminding them about some elements of a healthy lifestyle. \\
   	\hline
\textbf{Danger of entering a "search bubble"} (without the text interaction with a lot of context sometimes it is hard to find exactly what we are looking for, or to remember what it was). & A companion screen could mitigate this effect, if the user could glance at it and request to be read a specific result. \\
   	\hline

\end{tabular}
\end{margintable}

In our research we identified the main benefits our group of older adults saw in the Voice User Interface and Smart Home solutions, as well as the key barriers to their implementation and use. At the same time we would like to point out the limitations to our study, as the participants were technology-conscious older adults in Poland, however, as one of them (P6) put it "there are some categories of older people who are interested in smartphones, but there are some who spend their time with a crochet, just as there are different categories among young people." As such, some of our insights may extend to other groups, therefore to inform the design of VUI-based solutions, we present our preliminary findings regarding the benefits in Table \ref{benefits} and barriers in Table \ref{barriers}.

Moreover, apart from identifying key barriers and benefits, we have analyzed our participants' queries, comments and interaction with the VA in order to explore its nature and introduce some preliminary categories to take into account when designing VA functionalities, such as their role as everyday duties assistants, helping with setting alarms, shopping lists, weather and traffic information (P5), as well as leisure assistants including such specific applications as listening to audio books (P1), language learning (P6), telling a jokes (P7)) and posing SH safety assistants or caretakers able to turn off the light or the stove (P3).

\section{Conclusions and future work}
Our exploratory qualitative study allowed us to draw important preliminary conclusions relating to the needs of older adults, the benefits and barriers of using VA technology as well as the different possible applications of VA combined with SHT.

First, we identified a number of reasons for which VA interfaces combined with IoT are well adjusted to the needs of many older adults, both cognitive, including their need to understand the technology and take control of it, and physical, accounting for accessibility and convenience. We concluded that this technology is very promising as it has the potential to empower some groups of older adults.

Second, we enumerated the main benefits our group of older adults saw in the Voice User Interface and Smart Home solutions, as well as the key barriers to their implementation and use. These we then matched with relevant improvement suggestions and solutions or comments which ought to be further discussed and explored. 

Third, we find that our group of older adults could naturally identify various already available applications of VA with SHT and accept them as generally useful, as well as list multiple additional applications that could spare them considerable effort and free up their time. 

Therefore, we think that this study is an important voice in the debate on the various applications of voice-powered interaction to meet the needs of some older adults. At the same time we would like to point out the limitations to our study, as the participants were technology-conscious older adults in Poland. Thus, further research is required to verify the identified preliminary barriers and benefits as well as to explore the insights gathered and to investigate this solution with different potential user groups to verify which insights may be group specific and which are general. 


\begin{sidebar}
\section{Acknowledgments}
We would like to thank older adults from our Living Lab, those affiliated with Kobo Association who participated in this study.
\end{sidebar}

\bibliography{bibliography}


\begin{thebibliography}{14}


\ifx \showCODEN    \undefined \def \showCODEN     #1{\unskip}     \fi
\ifx \showDOI      \undefined \def \showDOI       #1{#1}\fi
\ifx \showISBNx    \undefined \def \showISBNx     #1{\unskip}     \fi
\ifx \showISBNxiii \undefined \def \showISBNxiii  #1{\unskip}     \fi
\ifx \showISSN     \undefined \def \showISSN      #1{\unskip}     \fi
\ifx \showLCCN     \undefined \def \showLCCN      #1{\unskip}     \fi
\ifx \shownote     \undefined \def \shownote      #1{#1}          \fi
\ifx \showarticletitle \undefined \def \showarticletitle #1{#1}   \fi
\ifx \showURL      \undefined \def \showURL       {\relax}        \fi
\providecommand\bibfield[2]{#2}
\providecommand\bibinfo[2]{#2}
\providecommand\natexlab[1]{#1}
\providecommand\showeprint[2][]{arXiv:#2}

\bibitem[\protect\citeauthoryear{Aula}{Aula}{2004}]%
        {aula_learning_2004}
\bibfield{author}{\bibinfo{person}{Anne Aula}.}
  \bibinfo{year}{2004}\natexlab{}.
\newblock \showarticletitle{Learning to use computers at a later age}.
\newblock In \bibinfo{booktitle}{\emph{{HCI} and the {Older} {Population}}}.
  \bibinfo{publisher}{Univ. of Glasgow}, \bibinfo{address}{Leeds, UK}.
\newblock


\bibitem[\protect\citeauthoryear{Boulton-Lewis, Buys, Lovie-Kitchin, Barnett,
  and David}{Boulton-Lewis et~al\mbox{.}}{2007}]%
        {boulton2007ageing}
\bibfield{author}{\bibinfo{person}{Gillian~M Boulton-Lewis},
  \bibinfo{person}{Laurie Buys}, \bibinfo{person}{Jan Lovie-Kitchin},
  \bibinfo{person}{Karen Barnett}, {and} \bibinfo{person}{L~Nikki David}.}
  \bibinfo{year}{2007}\natexlab{}.
\newblock \showarticletitle{Ageing, learning, and computer technology in
  Australia}.
\newblock \bibinfo{journal}{\emph{Educational Gerontology}}
  \bibinfo{volume}{33}, \bibinfo{number}{3} (\bibinfo{year}{2007}),
  \bibinfo{pages}{253--270}.
\newblock


\bibitem[\protect\citeauthoryear{Corbett and Weber}{Corbett and Weber}{2016}]%
        {Corbett:2016:ISA:2935334.2935386}
\bibfield{author}{\bibinfo{person}{Eric Corbett} {and} \bibinfo{person}{Astrid
  Weber}.} \bibinfo{year}{2016}\natexlab{}.
\newblock \showarticletitle{What Can I Say?: Addressing User Experience
  Challenges of a Mobile Voice User Interface for Accessibility}. In
  \bibinfo{booktitle}{\emph{Proceedings of the 18th International Conference on
  Human-Computer Interaction with Mobile Devices and Services}}
  \emph{(\bibinfo{series}{MobileHCI '16})}. \bibinfo{publisher}{ACM},
  \bibinfo{address}{New York, NY, USA}, \bibinfo{pages}{72--82}.
\newblock
\showISBNx{978-1-4503-4408-1}
\urldef\tempurl%
\url{https://doi.org/10.1145/2935334.2935386}
\showDOI{\tempurl}


\bibitem[\protect\citeauthoryear{Djoub}{Djoub}{2013}]%
        {djoub_ict_2013}
\bibfield{author}{\bibinfo{person}{Zineb Djoub}.}
  \bibinfo{year}{2013}\natexlab{}.
\newblock \showarticletitle{{ICT} education and motivating elderly people}. In
  \bibinfo{booktitle}{\emph{Ariadna; cultura, educación y tecnología}},
  Vol.~\bibinfo{volume}{1}. \bibinfo{pages}{88--92}.
\newblock
\urldef\tempurl%
\url{https://doi.org/10.6035/Ariadna.2013.1.15}
\showDOI{\tempurl}


\bibitem[\protect\citeauthoryear{Kope{\'{c}}, Balcerzak, Nielek, Kowalik,
  Wierzbicki, and Casati}{Kope{\'{c}} et~al\mbox{.}}{2018}]%
        {Kopeć2018}
\bibfield{author}{\bibinfo{person}{Wies{\l}aw Kope{\'{c}}},
  \bibinfo{person}{Bart{\l}omiej Balcerzak}, \bibinfo{person}{Rados{\l}aw
  Nielek}, \bibinfo{person}{Grzegorz Kowalik}, \bibinfo{person}{Adam
  Wierzbicki}, {and} \bibinfo{person}{Fabio Casati}.}
  \bibinfo{year}{2018}\natexlab{}.
\newblock \showarticletitle{Older adults and hackathons: a qualitative study}.
\newblock \bibinfo{journal}{\emph{Empirical Software Engineering}}
  \bibinfo{volume}{23}, \bibinfo{number}{4} (\bibinfo{date}{01 Aug}
  \bibinfo{year}{2018}), \bibinfo{pages}{1895--1930}.
\newblock
\showISSN{1573-7616}
\urldef\tempurl%
\url{https://doi.org/10.1007/s10664-017-9565-6}
\showDOI{\tempurl}


\bibitem[\protect\citeauthoryear{Kope{\'c}, Nielek, and Wierzbicki}{Kope{\'c}
  et~al\mbox{.}}{2018}]%
        {kopec2017spiral}
\bibfield{author}{\bibinfo{person}{Wies{\l}aw Kope{\'c}},
  \bibinfo{person}{Rados{\l}aw Nielek}, {and} \bibinfo{person}{Adam
  Wierzbicki}.} \bibinfo{year}{2018}\natexlab{}.
\newblock \showarticletitle{Guidelines Towards Better Participation of Older
  Adults in Software Development Processes using a new SPIRAL Method and
  Participatory Approach}. In \bibinfo{booktitle}{\emph{Proceedings of the
  CHASE'18: International Workshop on Cooperative and Human Aspects of
  Software}} \emph{(\bibinfo{series}{ICSE '18})}. \bibinfo{publisher}{ACM},
  \bibinfo{address}{New York, NY, USA}.
\newblock
\urldef\tempurl%
\url{https://doi.org/10.1145/3195836.3195840}
\showDOI{\tempurl}


\bibitem[\protect\citeauthoryear{Kope\'{c}, Skorupska, Jaskulska, Abramczuk,
  Nielek, and Wierzbicki}{Kope\'{c} et~al\mbox{.}}{2017}]%
        {kopec2017living}
\bibfield{author}{\bibinfo{person}{Wies{\l}aw Kope\'{c}},
  \bibinfo{person}{Kinga Skorupska}, \bibinfo{person}{Anna Jaskulska},
  \bibinfo{person}{Katarzyna Abramczuk}, \bibinfo{person}{Radoslaw Nielek},
  {and} \bibinfo{person}{Adam Wierzbicki}.} \bibinfo{year}{2017}\natexlab{}.
\newblock \showarticletitle{LivingLab PJAIT: Towards Better Urban Participation
  of Seniors}. In \bibinfo{booktitle}{\emph{Proceedings of the International
  Conference on Web Intelligence}} \emph{(\bibinfo{series}{WI '17})}.
  \bibinfo{publisher}{ACM}, \bibinfo{address}{New York, NY, USA},
  \bibinfo{pages}{1085--1092}.
\newblock
\showISBNx{978-1-4503-4951-2}
\urldef\tempurl%
\url{https://doi.org/10.1145/3106426.3109040}
\showDOI{\tempurl}


\bibitem[\protect\citeauthoryear{Naumanen and Tukiainen}{Naumanen and
  Tukiainen}{2008}]%
        {naumanen_practices_2008}
\bibfield{author}{\bibinfo{person}{Minnamari Naumanen} {and}
  \bibinfo{person}{Markku Tukiainen}.} \bibinfo{year}{2008}\natexlab{}.
\newblock \showarticletitle{Practices in old age {ICT} education}. In
  \bibinfo{booktitle}{\emph{{IADIS} {International} {Conference} on {Cognition}
  and {Exploratory} {Learning} in {Digital} {Age}}}. \bibinfo{pages}{261--269}.
\newblock


\bibitem[\protect\citeauthoryear{Pyae and Joelsson}{Pyae and Joelsson}{2018}]%
        {Pyae:2018:IUU:3236112.3236130}
\bibfield{author}{\bibinfo{person}{Aung Pyae} {and} \bibinfo{person}{Tapani~N.
  Joelsson}.} \bibinfo{year}{2018}\natexlab{}.
\newblock \showarticletitle{Investigating the Usability and User Experiences of
  Voice User Interface: A Case of Google Home Smart Speaker}. In
  \bibinfo{booktitle}{\emph{Proceedings of the 20th International Conference on
  Human-Computer Interaction with Mobile Devices and Services Adjunct}}
  \emph{(\bibinfo{series}{MobileHCI '18})}. \bibinfo{publisher}{ACM},
  \bibinfo{address}{New York, NY, USA}, \bibinfo{pages}{127--131}.
\newblock
\showISBNx{978-1-4503-5941-2}
\urldef\tempurl%
\url{https://doi.org/10.1145/3236112.3236130}
\showDOI{\tempurl}


\bibitem[\protect\citeauthoryear{Sandhu, Damodaran, and Ramondt}{Sandhu
  et~al\mbox{.}}{2013}]%
        {sandhu2013ict}
\bibfield{author}{\bibinfo{person}{Jatinder Sandhu}, \bibinfo{person}{Leela
  Damodaran}, {and} \bibinfo{person}{Leonie Ramondt}.}
  \bibinfo{year}{2013}\natexlab{}.
\newblock \showarticletitle{ICT Skills Acquisition by Older People: Motivations
  for learning and barriers to progression}.
\newblock \bibinfo{journal}{\emph{International Journal of Education and
  Ageing}} \bibinfo{volume}{3}, \bibinfo{number}{1} (\bibinfo{year}{2013}),
  \bibinfo{pages}{25--42}.
\newblock


\bibitem[\protect\citeauthoryear{Skorupska, N\'{u}\~{n}ez, Kope{\'c}, and
  Nielek}{Skorupska et~al\mbox{.}}{2018}]%
        {skorupska2018smarttv}
\bibfield{author}{\bibinfo{person}{Kinga Skorupska}, \bibinfo{person}{Manuel
  N\'{u}\~{n}ez}, \bibinfo{person}{Wies{\l}aw Kope{\'c}}, {and}
  \bibinfo{person}{Radoslaw Nielek}.} \bibinfo{year}{2018}\natexlab{}.
\newblock \showarticletitle{Older Adults and Crowdsourcing: Android TV App for
  Evaluating TEDx Subtitle Quality}. In \bibinfo{booktitle}{\emph{Proceedings
  of the 2018 ACM Conference on Computer Supported Cooperative Work and Social
  Computing}}. ACM, \bibinfo{pages}{286--296}.
\newblock


\bibitem[\protect\citeauthoryear{Vacher, Caffiau, Portet, Meillon, Roux, Elias,
  Lecouteux, and Chahuara}{Vacher et~al\mbox{.}}{2015}]%
        {Vacher:2015:ECV:2785580.2738047}
\bibfield{author}{\bibinfo{person}{Michel Vacher}, \bibinfo{person}{Sybille
  Caffiau}, \bibinfo{person}{Fran\c{c}ois Portet}, \bibinfo{person}{Brigitte
  Meillon}, \bibinfo{person}{Camille Roux}, \bibinfo{person}{Elena Elias},
  \bibinfo{person}{Benjamin Lecouteux}, {and} \bibinfo{person}{Pedro
  Chahuara}.} \bibinfo{year}{2015}\natexlab{}.
\newblock \showarticletitle{Evaluation of a Context-Aware Voice Interface for
  Ambient Assisted Living: Qualitative User Study vs. Quantitative System
  Evaluation}.
\newblock \bibinfo{journal}{\emph{ACM Trans. Access. Comput.}}
  \bibinfo{volume}{7}, \bibinfo{number}{2}, Article \bibinfo{articleno}{5}
  (\bibinfo{date}{May} \bibinfo{year}{2015}), \bibinfo{numpages}{36}~pages.
\newblock
\showISSN{1936-7228}
\urldef\tempurl%
\url{https://doi.org/10.1145/2738047}
\showDOI{\tempurl}


\bibitem[\protect\citeauthoryear{Vines, Pritchard, Wright, Olivier, and
  Brittain}{Vines et~al\mbox{.}}{2015}]%
        {articleageoldproblem}
\bibfield{author}{\bibinfo{person}{John Vines}, \bibinfo{person}{Gary
  Pritchard}, \bibinfo{person}{Peter Wright}, \bibinfo{person}{Patrick
  Olivier}, {and} \bibinfo{person}{Katie Brittain}.}
  \bibinfo{year}{2015}\natexlab{}.
\newblock \showarticletitle{An Age-Old Problem: Examining the Discourses of
  Ageing in HCI and Strategies for Future Research}.
\newblock \bibinfo{journal}{\emph{ACM Trans. Comput.-Hum. Interact.}}
  \bibinfo{volume}{22}, \bibinfo{number}{1}, Article \bibinfo{articleno}{2}
  (\bibinfo{date}{Feb.} \bibinfo{year}{2015}), \bibinfo{numpages}{27}~pages.
\newblock
\showISSN{1073-0516}
\urldef\tempurl%
\url{https://doi.org/10.1145/2696867}
\showDOI{\tempurl}


\bibitem[\protect\citeauthoryear{von Helversen, Abramczuk, Kope{\'c}, and
  Nielek}{von Helversen et~al\mbox{.}}{2018}]%
        {von2018influence}
\bibfield{author}{\bibinfo{person}{Bettina von Helversen},
  \bibinfo{person}{Katarzyna Abramczuk}, \bibinfo{person}{Wies{\l}aw
  Kope{\'c}}, {and} \bibinfo{person}{Radoslaw Nielek}.}
  \bibinfo{year}{2018}\natexlab{}.
\newblock \showarticletitle{Influence of Consumer Reviews on Online Purchasing
  Decisions in Older and Younger Adults}.
\newblock \bibinfo{journal}{\emph{Decision Support Systems}}
  (\bibinfo{year}{2018}).
\newblock


\end{thebibliography}

\bibliographystyle{ACM-Reference-Format}

\end{document}